# Efficient Methods for Handling Long-Range Forces in Particle-Particle Simulations

Hans Fangohr[†][‡] fangohr@soton.ac.uk

Andrew R. Price[‡] arp97r@ecs.soton.ac.uk

Simon J. Cox[‡] sc@ecs.soton.ac.uk

Peter A.J. de Groot[†] pajdeg@phys.soton.ac.uk

Geoffrey J. Daniell[†] gjd@phys.soton.ac.uk

Ken S. Thomas[†] kst@ecs.soton.ac.uk

[†]*Department of Physics and Astronomy*

[‡]*Department of Electronics and Computer Science*

*University of Southampton, Southampton, SO17 1BJ, UK*

# ABSTRACT

A number of problems arise when long-range forces, such as those governed by Bessel functions, are used in particle-particle simulations. If a simple cut-off for the interaction is used, the system may find an equilibrium configuration at zero temperature that is not a regular lattice yet has an energy lower than the theoretically predicted minimum for the physical system. We demonstrate two methods to overcome these problems in Monte Carlo and molecular dynamics simulations. The first uses a smoothed potential to truncate the interaction in a single unit cell: this is appropriate for phenomenological characterisations, but may be applied to any potential. The second is a new method for summing the unmodified potential in an infinitely tiled periodic system, which is in excess of 20,000 times faster than previous naïve methods which add periodic images in shells of increasing radius: this is suitable for quantitative studies. Finally we show that numerical experiments which do not handle the long-range force carefully may give misleading results: both of our proposed methods overcome these problems.

## 1. INTRODUCTION

Considerable effort has been invested in handling long-range forces for particle-particle simulations. The conventional cut-off approach truncates the potential in a single unit cell for separations greater than half the system dimension. In general it is better to sum the potential over a number of repeats of the unit cell. Infinite summation methods include the Ewald summation [1, 2, 3], multipole methods [4], lattice summation methods [5], the Lekner summation method [6, 7] and a novel method for logarithmic interactions [8]. In this paper we review some of the problems which can occur when the potential is naïvely truncated, which have not previously been widely reported in the literature. We then derive two methods which overcome these problems. The first is suitable for phenomenological studies of systems and smooths the potential within a single unit cell. The second is a new real-space summation method appropriate for potentials governed by Bessel functions. This provides a speed-up of at least 20,000 compared to the current method of summing in a series of shells of increasing radius [9].

In section 2 we introduce our model system, which is a simulation of a layered superconductor. We discuss the problems which arise with cutting off this potential in a single unit cell in section 3, and give a simple method of smoothing the potential which overcomes these problems in section 4. In section 5 we consider an infinitely tiled periodic system and derive our new summation method. Section 6 describes a simulation of shearing a superconductor lattice using our new methods and constrasts it with the results obtained when the potential is cut-off. We draw our conclusions in section 7.

## 2. MODEL SYSTEM

We will consider the long-range forces which arise in the simulation of pancake vortices in layered high-temperature superconductors [10]. The potential is governed by [9, 11, 12, 13, 14]:

$$\frac{U(r)}{c} = K_0\left(\frac{r}{\lambda}\right),$$

(1)

where $\lambda$ is the penetration depth of the magnetic field, $r$ is the distance between the particles and $c$ is a constant. This may be approximated as

$$\frac{U(r)}{c} = \begin{cases} \left(\frac{\pi\lambda}{2r}\right)^{1/2}\exp\left(-\frac{r}{\lambda}\right) & r \to \infty \\ \ln\left(\frac{\lambda}{r}\right) + 0.12 & r << \lambda \end{cases}.$$

(2)

Since $\lambda$ can be several orders of magnitude larger than $r$ [9], the $K_0$ potential has a very long range character. It is therefore necessary to either (i) only consider the interaction inside a single unit cell which contains a large number of particles, or (ii) sum the interaction over period repeats of the unit cell.

Our findings are also of relevance to the simulation of other systems governed by long-range forces such as the interaction of electrically charged rods [8]. We will show results for Monte Carlo and Molecular Dynamics simulations where the two-dimensional unit cell geometry can be chosen to be a rectangle, a parallelogram or a hexagon. In all cases periodic boundary conditions are employed.

## 3. CUT-OFF POTENTIAL

The standard approach is to cut-off the potential to be constant outside a circle of radius equal to $\min(L_x / 2,\ L_y / 2)$, where $L_x$ and $L_y$ are the lengths of the sides of the unit cell. Since the force is the gradient of the potential, it is zero outside the cut-off radius. We then define the distance between particles, $r$, to be the minimum image distance [3].

In figure 1 the real force dependence $F(r)$ is compared to that for a simulation system with a simple geometrical cut-off. For vortices in superconductors, Abrikosov [15] demonstrated theoretically that the lowest energy configuration for an infinite lattice is the hexagonal lattice, or so-called Abrikosov lattice, with an associated Abrikosov lattice energy. However, when using a sharp cut-off in our simulations we find many configurations with energies lower than the Abrikosov lattice energy.

Figure 2 shows the results from a Molecular Dynamics simulation of a small number of particles in which the temperature in the system is cycled from 0K to half the melting temperature of the vortex solid and is then returned to 0K. The temperature is introduced via a stochastic noise term. The Delaunay triangulation of the vortex configuration at the end of the simulation in figure 2 is elastically deformed. Detailed examination of the triangulation shows that the elastic deformations arise due to particles gathering on the boundaries of the cut-off circles. In this position they minimise their contribution to the energy (or force) in the system. This gives rise to the "wavy lines" visible in figure 2, with a curvature characterised by the cut-off radius. To demonstrate this, we have shown the cut-off circles corresponding to two of the particles. The wavy lines are less evident in larger systems, since their curvature is inversely proportional to the cut-off radius.

If the system is heated above its melting temperature and then annealed slowly the final equilibrium state (i) has an energy lower than the Abrikosov energy, and (ii) contains topological defects. A topological defect is a particle which does not have six nearest neighbours in the Delaunay triangulation. We have repeated these results for Molecular Dynamics and Monte Carlo simulations with up to 2000 particles. The result in figure 3 for a Monte Carlo simulation of a system annealed from a liquid state exhibits low energy and contains defects. We have verified that our results are independent of the geometry of the unit cell (rectangular, parallelogram, or hexagonal).

These problems are clearly artificial, and are caused by imposing a sharp cut-off on the very long range nature of the interaction. Since the penetration depth, $\lambda$, is generally much larger than the lattice spacing it would require systems with several hundred thousand particles before the effects of this finite size problem began to become less significant. Methods to deal with such large systems with the Bessel function interaction potential are currently being developed [16].

In studies of high temperature superconductors, interest has recently developed in the formation of topologically ordered states which exhibit quasi-long range translational order: the so-called Bragg glass. These states occur when the vortices are weakly pinned and have been investigated both theoretically and experimentally [17, 18]. Other studies have focussed on the structural properties of the dynamics of vortex systems [19, 20]. In both cases it is important that the ground state for an unpinned system should be a hexagonal lattice without topological defects. Furthermore for the calculation of numerical phase diagrams as a function of disordering pinning, it is vital that the disorder is not introduced by the model itself.

We therefore propose two methods which avoid the problems described above. The first involves modifying the potential near to the cut-off, and allows qualitative simulation of small systems using only a single unit cell. The second is a new fast summation method to allow the infinitely tiled periodic system to be considered and allow quantitative simulations to be performed.

## 4. SMOOTHED POTENTIAL

In figure 4 (left) we show the force field experienced by a vortex due to its surrounding particles in a hexagonal configuration. The discontinuities are caused by the artificial step in the force function shown in figure 1. It is natural to introduce a smoothed potential, which reduces the force smoothly to zero over a region from $r_{\text{fade}}$ to $r_{\text{cut-off}}$, and we impose $C^1$ continuity of the force at $r = r_{\text{fade}}$ and $r = r_{\text{cut-off}}$. The smoothed potential is shown in figure 1, with the resulting smooth force field in figure 4 (right). The smoothing distance $r_{\text{cut-off}} - r_{\text{fade}}$ is a free parameter which should be kept as small as possible to maintain the original force over the largest possible range. Numerical experiments show that three lattices spacings is sufficient. Figure 5 shows the results of a Monte Carlo simulation using a similarly smoothed energy. Simulations using this modified potential do not find configurations below the Abrikosov energy and topological defects only occur when the system is annealed very rapidly.

The interpretation is that due to the slow force change at the cut-off (enforced by the derivative being zero) a particle pair separated by a distance of roughly $r_{\text{cut-off}}$ experiences continuous and small changes in force if their positions are perturbed. This is in contrast to the large discontinuous fluctuations, which can enable the system to discover configurations with energies less than the Abrikosov energy. We have also used interpolating polynomials of higher order and an exponential function in the smoothing region: in all cases the system does not discover energy states below the Abrikosov energy.

It is important to consider whether the modification of the original force with the smooth cut-off affects the system's behaviour. Using a cut-off to the long-range interaction is a major change of the long-range interaction. However, introducing the smoothing distance and altering the force in the region between $r_{\text{fade}}$ and $r_{\text{cut-off}}$ cannot be worse than using a slightly smaller system with $r'_{\text{cut-off}} = r_{\text{fade}}$. The enormous advantage of using a smooth cut-off is that the structural properties of the system can be simulated correctly and that the lowest energy configuration is identical to the theoretical ground state. For studies of the dynamics of vortices, recent results show that the precise details of the long-range particle interaction are not crucial [13]. We therefore recommend the smoothed potential for phenomenological characterisation of superconductors.

## 5. FAST INFINITE SUMMATION

An alternative approach to modifying the potential is to sum the potential function over periodic repeats of the unit cell, which provides the best representation of the system given only a finite number of particles. We write the potential (1) in the form: [9]

$$\frac{U(r)}{c} = K_0^* \left( \frac{|r|}{\lambda} \right) = \sum_{m_x, m_y} K_0 \left( \frac{r + L_x m_x \hat{x} + L_y m_y \hat{y}}{\lambda} \right),$$ (3)

where $m_x$ and $m_y$ are integers and $L_x$ and $L_y$ are the lengths of the edges of the simulation cell. This is truncated such that $m_x^2 + m_y^2 \le N_m^2$; we sum the potential in shells of increasing radius, $N_m$, until it has converged. Following Ryu *et al* [9] we will use a value for the penetration depth, $\lambda$, at 0K of 7700 Å for $Mo_{77}Ge_{23}$. We will return to the temperature dependence of $\lambda$ later. In figure 6, we show the exponentially fast convergence of the energy between two particles in a simulation of 300 vortices in the Abrikosov lattice state as more image cells are included. We also show the time taken to perform this calculation on a 450 MHz Pentium II using Compaq (Digital) Visual Fortran under Windows NT 4.0. For the particle-particle energy to converge to a relative error better than $1 \times 10^{-8}$ requires $N_m \sim 300$, which takes $\sim 300,000$ calls to the $K_0$ function. This ensures that the total system energy is accurate to better than 0.01%.

We now derive a new method to perform this infinite summation. In figure 7 we have:

$$Z^2 = (m_x L_x)^2 + (m_y L_y)^2$$
$$z^2 = (x_i - x_j)^2 + (y_i - y_j)^2$$
$$\theta = \tan^{-1} \left( \frac{x_i - x_j}{y_i - y_j} \right) + \frac{\pi}{2}$$ (4)
$$\varphi = \tan^{-1} \left( \frac{m_y L_y}{m_x L_x} \right),$$

which yields

$$\phi = \theta + \varphi$$
$$w^2 = Z^2 + z^2 - 2zZ \cos(\phi).$$ (5)

We may use the Gegenbauer addition formulae [21] to write

$$K_0 \left( \frac{w}{\lambda} \right) = \sum_{k=-\infty}^{\infty} K_k \left( \frac{Z}{\lambda} \right) I_k \left( \frac{z}{\lambda} \right) \cos(k\phi)$$ (6)

for the energy between a particle $i$ and one of the periodic images of $j$, where $I_k$ and $K_k$ are modified Bessel functions. This formula requires $z \le Z$, which is automatically satisfied since $z$ is the minimum image distance between $i$ and $j$. We can therefore write the total energy (3) of two particles $i$ and $j$ summed over all periodic images in the form:

$$K_0^* \left( \frac{|r|}{\lambda} \right) = K_0^* \left( \frac{w}{\lambda} \right) = K_0 \left( \frac{z}{\lambda} \right) + \sum_{\substack{m_x, m_y \\ \text{not } m_x = m_y = 0}} \sum_{k=-\infty}^{\infty} K_k \left( \frac{Z}{\lambda} \right) I_k \left( \frac{z}{\lambda} \right) \cos(k\phi)$$ (7)

where the case $m_x = m_y = 0$, for which z $\Leftrightarrow$ Z, is the contribution to the energy from the unit cell which must be explicitly included as a separate term. Further re-arrangement and use of (5) gives

$$K_0^* \left( \frac{w}{\lambda} \right) = K_0 \left( \frac{z}{\lambda} \right) + \sum_{k=-\infty}^{\infty} I_k \left( \frac{z}{\lambda} \right) \left[ c_k \cos(k\theta) - s_k \sin(k\theta) \right]$$ (8)

where

$$c_k = \sum_{\substack{m_x, m_y \\ \text{not } m_x = m_y = 0}} K_k\left(\frac{Z}{\lambda}\right)\cos(k\varphi) \quad \text{and} \quad s_k = \sum_{\substack{m_x, m_y \\ \text{not } m_x = m_y = 0}} K_k\left(\frac{Z}{\lambda}\right)\sin(k\varphi) \tag{9}$$

Equations (8) and (9) have the remarkable property that the coefficients corresponding to the infinite summation over the periodic repeats of the unit cell can be pre-computed. This reduces the double summation in (3) to a single summation. Furthermore, due to the exponential convergence of the Gegenbauer addition formulae, the sum may be truncated at $k_{\text{trunc}} \sim 5 - 20$ terms. A further factor of two in performance can be obtained by using symmetry to convert the summation from $k = -\infty \ldots \infty$ to the range $k = 0 \ldots \infty$.

The form (8) closely resembles a Fourier type summation method, yet the whole calculation proceeds in real space in contrast to the Ewald summation method [22]. Our proposed method couples directly to a multipole method for computing the interaction energy inside the unit cell in O($N$) time [16], which is based on the Gegenbauer addition formulae, rather than a Taylor series expansion. Our O($N$) method provides further speedup when there are more than ~1200 particles in the unit cell. This is analogous to the method described in [5], which couples a lattice summation method with a multipole method based on Taylor series. It is certainly not appropriate to use the method proposed in [8], which sums a genuinely logarithmic potential over infinite repeats of the unit cell, since the logarithmic approximation to the $K_0$ potential is only valid for small $r$ as shown in (2).

The convergence of the energy between two particles in the Abrikosov lattice is identical to the convergence shown in figure 6 as we add more terms to the calculation of the coefficients $c_k$ and $s_k$. We have chosen the case of two nearest neighbours, which yields the slowest convergence of (8) since $z$ takes its smallest value.

In a superconductor, $\lambda$ is a function of the temperature. For our model system ($Mo_{77}Ge_{23}$) $\lambda(T) = \lambda(0) / (1-T / T_c)^{1/2}$ [9], where $T_c = 5.63$K is the critical temperature at which the material loses its superconducting properties. Hence the coefficients $c_k$ and $s_k$ need to be re-computed at each temperature. As the temperature increases additional image cells need to be included in both (3) and the pre-computation (9). The crucial difference, however, between (3) and (8) is that the time taken to evaluate the energy using (8) remains constant once the coefficients are available, whereas the naïve summation requires considerable numbers of additional image cells to converge to the solution. In figure 8 we show the speedup of our method when computing the energy between two particles at a fixed accuracy of $1\times10^{-5}$ (relative to the energy computed to machine accuracy by either method). In all cases the resulting energies are shown to be identical to the stated accuracy. At 0K and using 5 terms in the truncation of (8), we have a speedup of 20,000 over the naïve summation method. This rises to 400,000 for temperatures approaching $T_c$. If the particle energy is required to be accurate to $1\times10^{-8}$, then, using 30 coefficients, the speedups are between 50,000 ($T = 0$K) and 1,000,000 (T ~ $T_c$).

Since the coefficients $c_k$ and $s_k$ depend on $\lambda$ (and hence temperature); the method may appear to be costly if the temperature is changed at every Molecular Dynamics or Monte Carlo step. We now discuss several ways to overcome this. Firstly, it is possible to perform simulations at a small number of temperatures and use the data from these to obtain information about the behaviour of the system as a continuous function of temperature [23, 24]. Thus improving the sophistication of the analysis of the results can reduce the number coefficients $c_k$ and $s_k$ which need to be pre-calculated. Secondly it is possible to compute the $c_k$ and $s_k$ at a small set of temperatures and use interpolation to derive their values at other temperatures. Finally, since only ~5-20 coefficients are needed, it is straightforward to compute once and store on disk the values of $c_k$ and $s_k$ for each temperature to be explored. These values will be re-used a large number of times in a typical set of numerical simulations.

We implement (8) using a recurrence relation [25] for the trigonometric terms and a vendor-optimised vector Bessel function. Goertzel's algorithm [26] could be employed for additional efficiency, though the improvement is likely to be marginal. The remarkable speedup obtained is due to the fixed work equivalent to roughly five calls to a Bessel function routine required for (8), compared to ~100,000 calls required for (3) (at 0K). The five calls are: two to initialise the Bessel recurrence, one to evaluate the contribution from the unit cell, and the equivalent of roughly a further two for the remaining trigonometric terms. Our infinite summation is correspondingly five times slower than using the smoothed potential in a single unit cell, which requires evaluation of a single Bessel function or a polynomial. This is confirmed by experiments. For simulations using the fast infinite lattice summation, results are similar to those of figure 5. The infinite lattice summation method is suitable for quantitative studies of superconductors.

## 6. RESULTS

In the previous sections we have demonstrated that the phenomenological potential and the infinitely summed potential ensure that the Abrikosov lattice is the minimum energy configuration for our system. We now show that the presence of dislocations, which also results from incorrect handling of the long-range potential, seriously affects study of the elastic properties of a lattice. For superconductors the structure of the lattice determines the static and dynamic properties of the vortex lattice. This is known from experimental [27, 28] and theoretical work [29]. The simulation potential should not introduce dislocations, since this will affect the onset of plasticity in the lattice which is directly related to characterising current-voltage behaviour, and thus to applications.

We have considered a simulation of shearing of a hexagonal lattice, which is a simplified version of the simulations required to perform current-voltage characterisations. Inset (a) in figure 9 shows a Delaunay triangulation for half the simulation cell demonstrating the experimental set-up: a shearing force is applied to the central row of particles marked by black points, and the particles marked by open circles are not allowed to move in the $x$-direction. The main diagram shows the resulting change in energy as a response to the shearing force. The upper part of the figure shows data for the smooth cut-off, with the lower part showing the results for the sharp cut-off. The smooth cut-off and the infinite lattice summation produce the expected behaviour: with increasing shear stress the energy increases. The slope of the energy-change as a function of the displacement characterises the shear elastic modulus of the crystal. Inset (b1) shows a triangulation of a system which has been slightly tilted by the applied force. In contrast, employing the sharp cut-off the energy decreases for applied shear stress, *i.e.* the material appears to collapse after applying a shearing force (inset b2)!

Insets (c1) and (c2) show the time evolution of the local hexatic order, $\Psi_6 = \frac{1}{n_{bond}} \left| \sum_k \exp(i6\theta_k) \right|$, where the sum runs over all bond angles $\theta_k$ in the Delaunay triangulation. Every 50,000 time steps the system starts as a hexagonal lattice ($\Psi_6 = 1$) and a new shearing force is applied for the next 50,000 time steps.

In (c1), which shows the smoothed potential, $\Psi_6$ decreases continuously until a static state is reached, reflecting the shearing of the system. The energy data is taken from these static states. In (c2) (sharp cut-off) $\Psi_6$ drops suddenly to a much smaller value, representing the sudden change to configurations similar to those shown in (b2). Thus, the mechanical properties of the lattice using a sharp cut-off are severely affected by the incorrect handling of the long-range potential: this would seriously affect numerical simulations aimed at studying elastic properties of superconductors. The smooth cut-off and the infinite lattice sum produce the correct physical behaviour and can be used in more complex numerical simulations for phenomenological (smoothed potential) or quantitative (infinite summation) study of the dynamic phase diagram of the superconductor lattice [30, 31].

## 7. CONCLUSIONS

For Monte Carlo and Molecular Dynamics simulations using long-range interactions subject to periodic boundary conditions, a sharp cut-off for the interaction energy (or force) can yield misleading results. We have considered the case of superconductors, in which the potential is governed by a Bessel function. Monte Carlo simulations are often used to study phase diagrams numerically and it is vital that the phase behaviour of the system is not affected by the model itself. We find that using a sharp cut-off the system can find irregular lattice configurations with an energy below the theoretical ground state of a regular hexagonal lattice. In Molecular Dynamics study of the dynamical phase diagram of the material can be dramatically affected by incorrect handling of the long-range potential.

We have presented two methods which overcome these problems. The first is suitable for phenomenological studies of systems and uses a smoothed potential, but still truncates the interaction over a single unit cell. Annealing a system governed by this modified potential yields a perfect hexagonal lattice which is the global energy minimum. This is the least computationally expensive option and is applicable to any potential. The second sums the interaction over the infinitely tiled unit cell and is suitable for quantitative system studies. Previous methods for performing this add the tiled images in a series of shells of increasing radius. We have shown that with the pre-computation of a set of Fourier type coefficients, the whole infinite summation can be computed using a summation which converges exponentially fast and results in a speedup of between 20,000 and 1,000,000 over the naïve summation, depending on the range of the interaction and the desired accuracy. The derivation of the summation proceeds in real space, and the results converge exactly to those obtained from other summation methods. This is roughly five times as slow as using the smoothed potential, but is the most accurate method for systems of finite size. We will report elsewhere on the results of systems we have studied using our methods [30, 31] and also on a method for evaluating the energy within the unit cell in O($N$) time [16].

## ACKNOWLEDGEMENTS

HF and ARP are grateful to EPSRC for studentships.

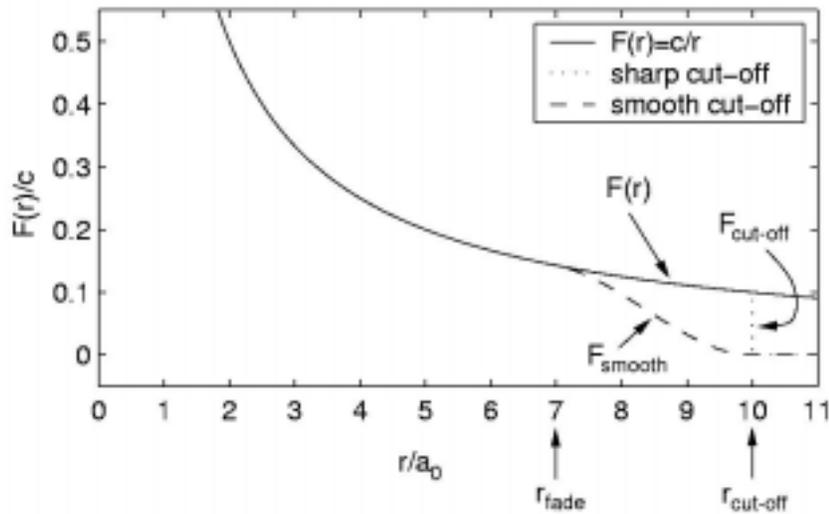

**Figure 1**

A long-range force (i) Full force (ii) Force cut-off at a distance $r_{cut-off}$ (iii) Smoothed force. Distances are measured in multiples of the ground-state lattice spacing.

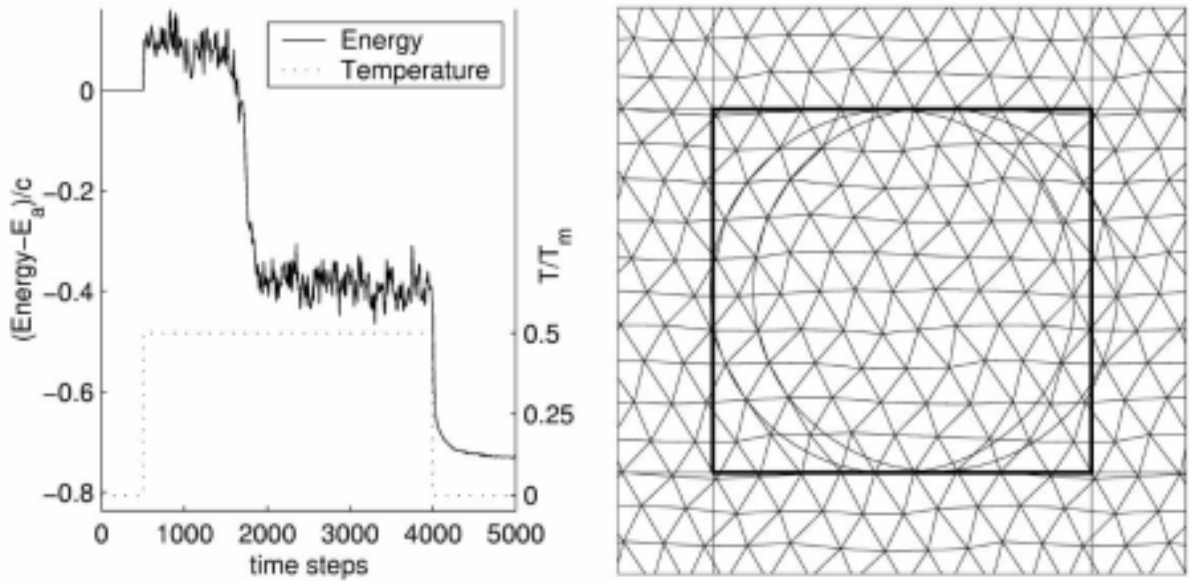

**Figure 2**

Left: Molecular Dynamics simulation of 90 particles using a cut-off potential, which start in a hexagonal configuration at 0K (with Abrikosov lattice energy, $E_a$), are heated to half their melting temperature ($T_m$) and then returned to 0K. Temperature is introduced via a stochastic noise term. The system finds a new configuration with energy lower than the energy of the regular lattice. Right: Delaunay triangulation of the final configuration of the particles at time step 5000. Two cut-off circles are shown to demonstrate that particles align along these circles.

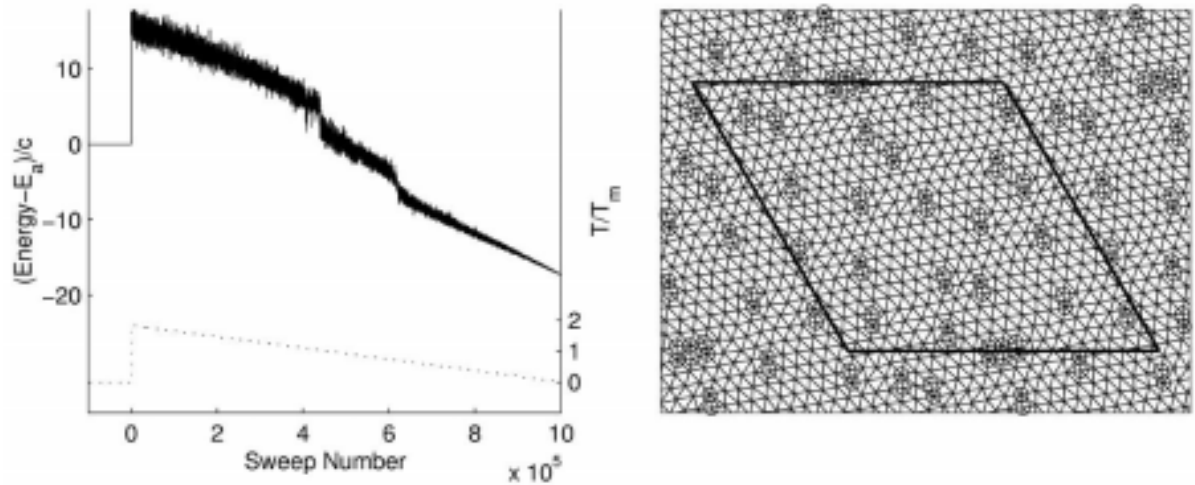

Figure 3

Monte Carlo simulation of 432 particles using a conventional cut-off potential. The system starts in a regular hexagonal Abrikosov lattice and is heated above its melting point to $\sim 2T_m$ then annealed slowly to zero temperature in steps of $0.01T_m$ each of 5000 sweeps. Left: The energy of the system drops below the Abrikosov lattice energy, $E_a$. Right: Delaunay triangulation of the final disordered configuration. The topological defects are circled.

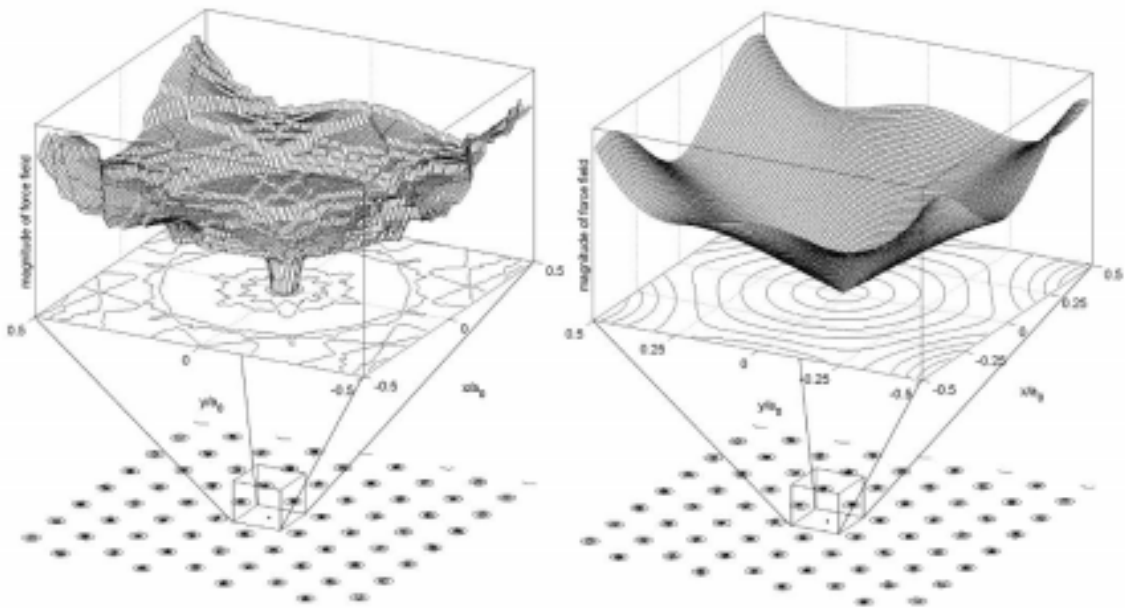

Figure 4

The magnitude of the force field a particle at position (0,0) experiences from a system of 418 particles using (left) the sharp cut-off and (right) the smooth cut-off. The effect of smoothing the potential is to remove the discontinuities in the force.

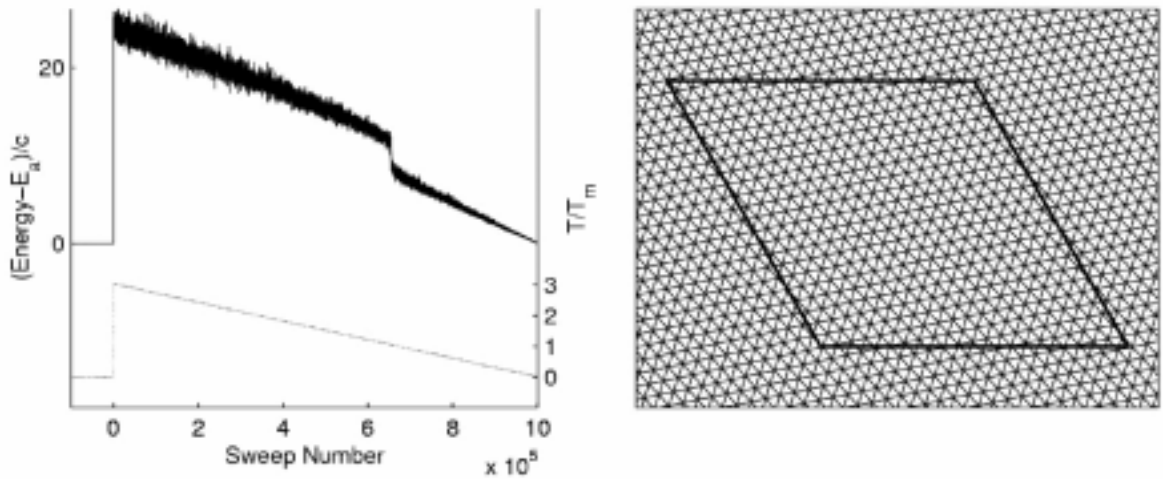

Figure 5

Monte Carlo simulation of 432 particles using a potential smoothed over three lattice spacings. Left: The energy of the system never drops below the Abrikosov lattice energy, $E_a$. Right: Delaunay triangulation of the final configuration shows the system has a hexagonal ground state.

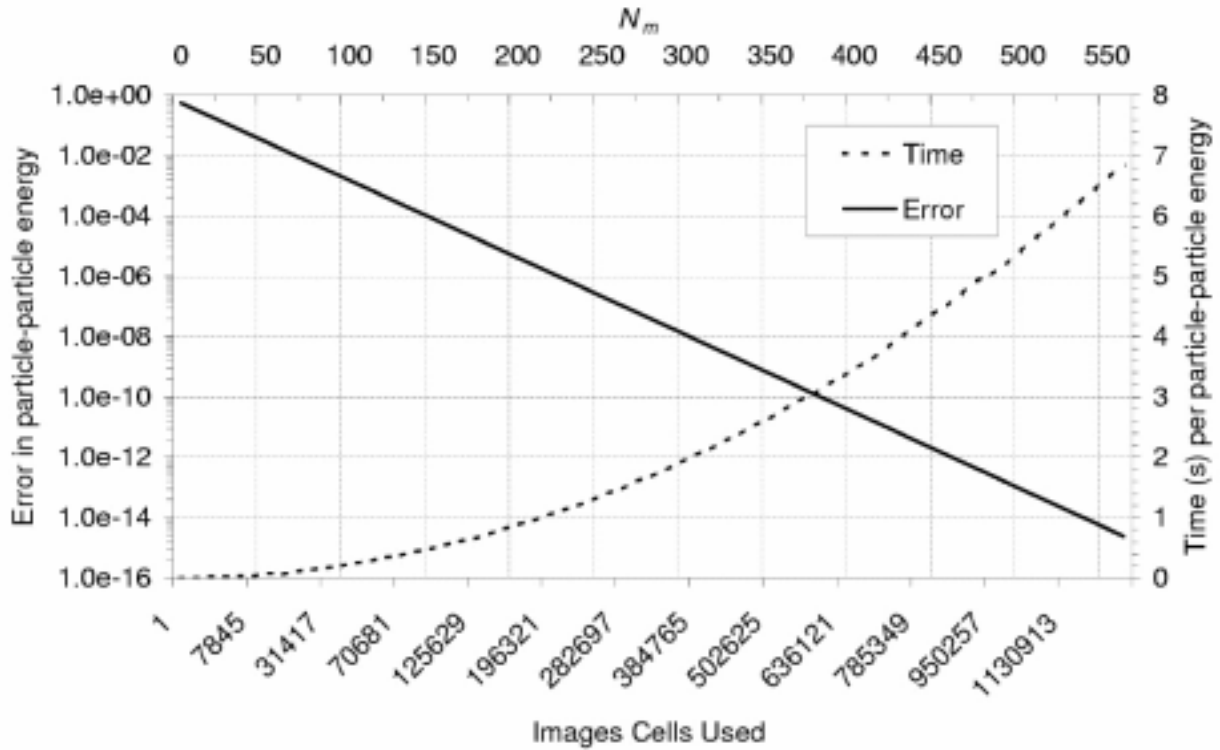

Figure 6

Fractional error $(E_\infty - E_n)/E_\infty$ and time taken to compute the energy $E_n$ between two particles separated by a single lattice spacing in an infinitely tiled periodic system when $n$ image cells are used. $E_\infty$ is estimated by allowing the summation to converge to machine accuracy.

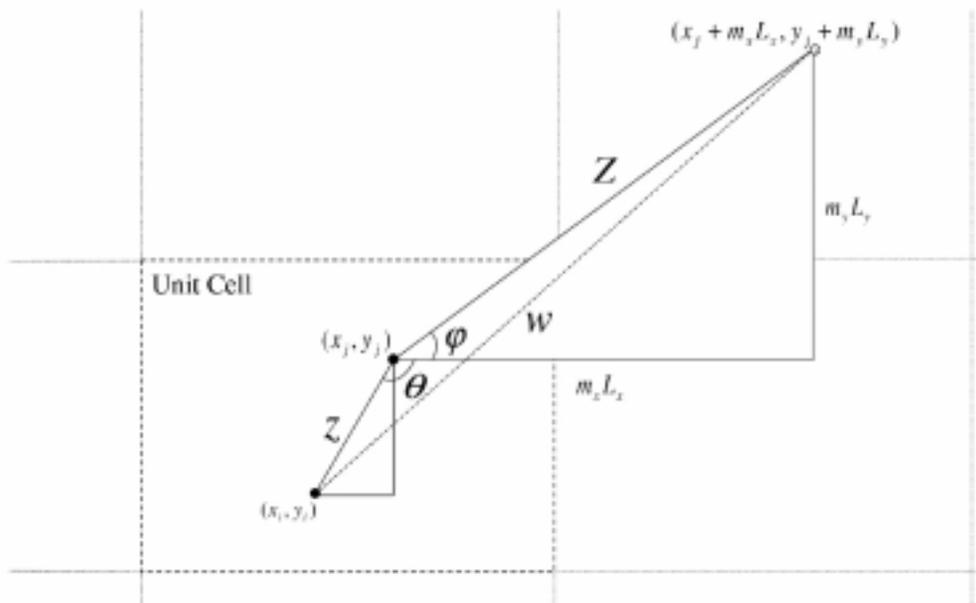

Figure 7

Two particles in a unit cell with infinite periodic repeats.

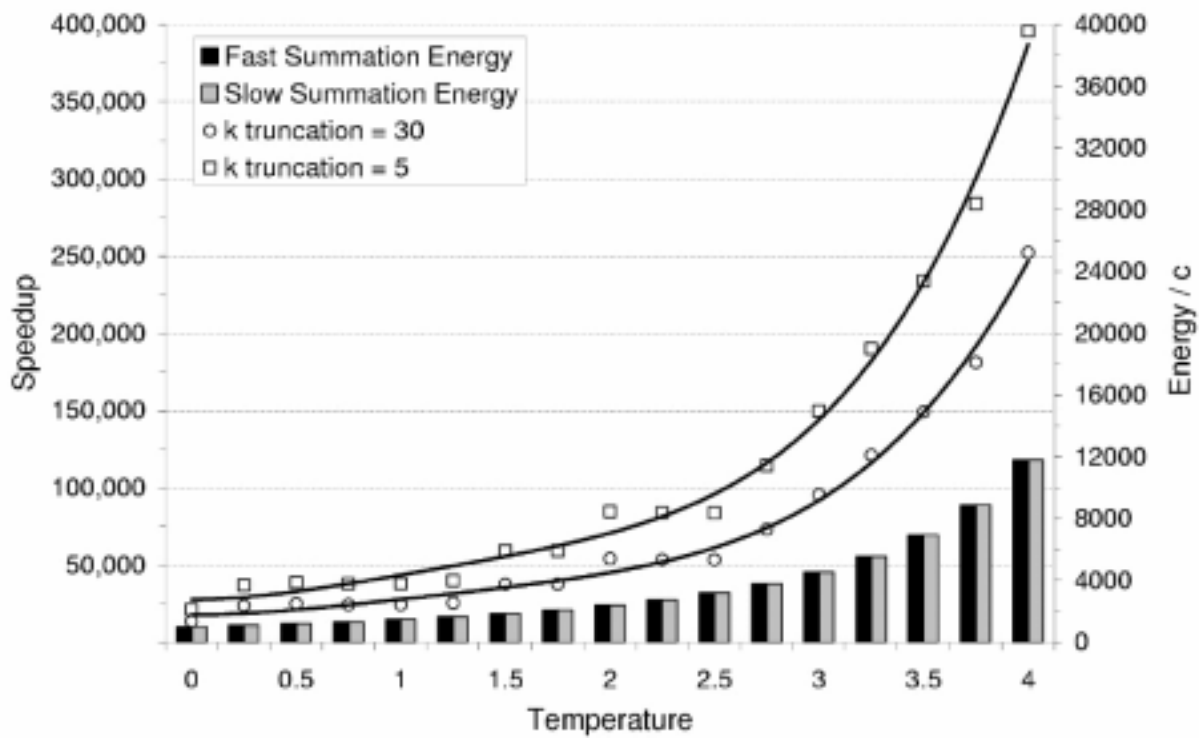

Figure 8

Speedup of fast infinite summation method over naïve implementation when the relative error in the energy between each pair of particles is fixed to be $1 \times 10^{-5}$: both methods yield identical results.

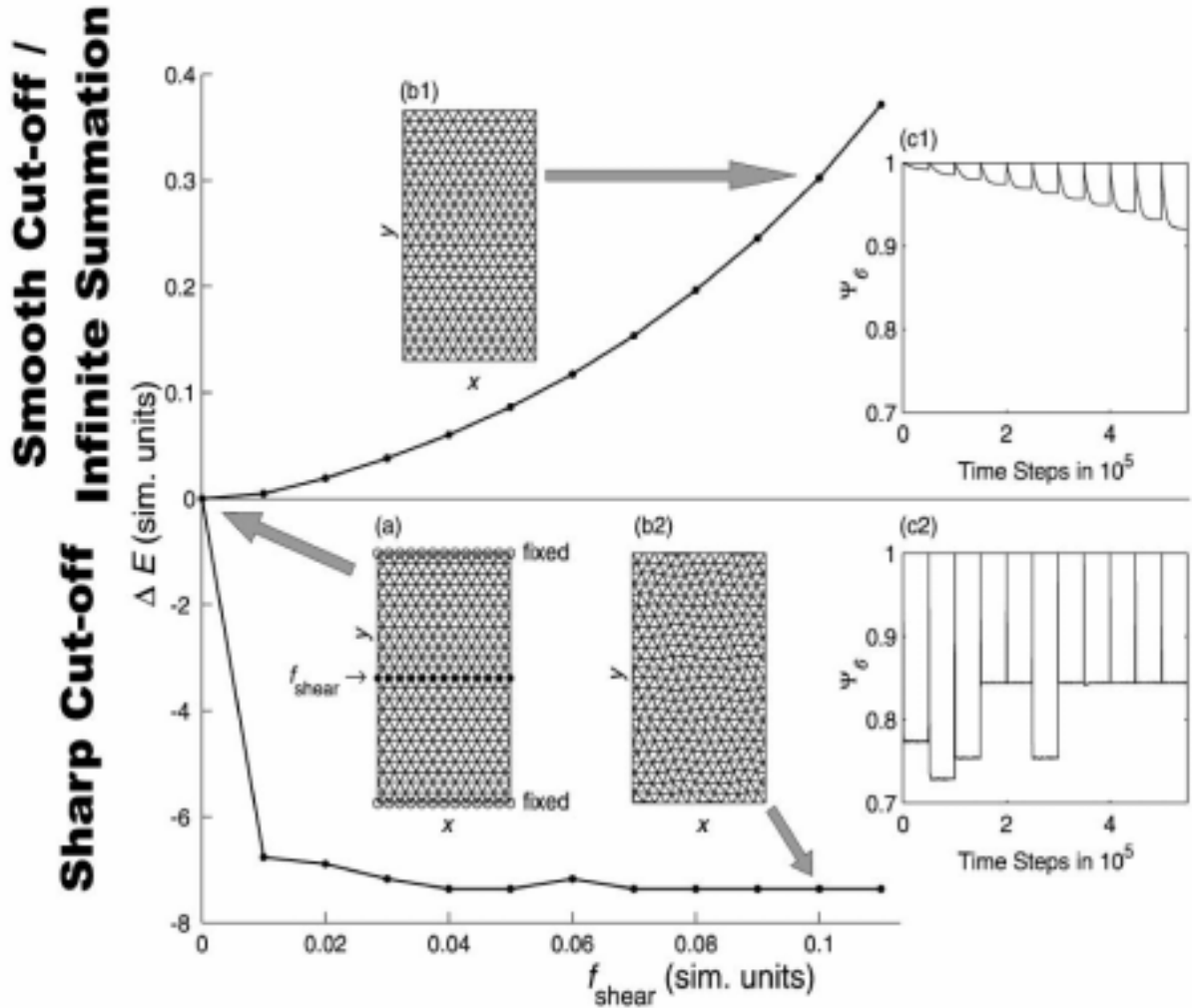

**Figure 9**

Change in energy, $\Delta E$, (in simulation units) as a function of a shearing force, $f_{shear}$, (in simulation units) for the smooth and the sharp cut-off. For the infinite lattice summation we obtain qualitatively similar results. Insets (a), (b1) and (b2) show different snap shots of vortex configurations. Insets (c1) and (c2) show the local hexagonal order, $\Psi_6$, as the experiment progresses (see text for details).